\documentstyle[11pt,aasms4,flushrt,epsf,rotate]{article}

\def\bt{\begin{tabbing}}
\def\et{\end{tabbing}}
\def\beq#1{\begin{equation}\label{#1}}
\def\eeq{\end{equation}}
\def\xhat{\widehat{\bf x}}
\def\khat{\widehat{\bf k}}

\def\spose#1{\hbox to 0pt{#1\hss}}
\def\simlt{\mathrel{\spose{\lower 3pt\hbox{$\mathchar"218$}}
     \raise 2.0pt\hbox{$\mathchar"13C$}}}
\def\simgt{\mathrel{\spose{\lower 3pt\hbox{$\mathchar"218$}}
     \raise 2.0pt\hbox{$\mathchar"13E$}}}      

\slugcomment{Published in {\it{ApJ}}, 448:477(1995).}

\begin{document}

 \title{CONSTRAINTS ON THE TOPOLOGY OF THE UNIVERSE FROM THE 2-YEAR COBE DATA}

\bigskip

\author{Ang\'{e}lica de Oliveira-Costa$^{1,2}$ and George F. Smoot$^{1}$}

\bigskip

\affil {$^1$Lawrence Berkeley Laboratory,  Space Sciences 
Laboratory and Center for Particle Astrophysics, Building 50-205, University 
of California, Berkeley, CA 94720; angelica@cosmos.lbl.gov}

\affil {$^2$Instituto Nacional de Pesquisas Espaciais (INPE), 
Astrophysics Division, S\~{a}o Jos\'{e} dos Campos, S\~{a}o Paulo 12227-010, 
Brazil.}

\begin{abstract}
The cosmic microwave background (CMB) is a unique probe of 
cosmological parameters and conditions. There is a connection 
between anisotropy in the
CMB and the topology of the Universe. Adopting a universe with the topology of 
a 3-Torus, or a universe where only harmonics of the fundamental mode are 
allowed, and using 2-years of COBE/DMR data, we obtain constraints on the 
topology of the Universe. Previous work constrained the topology using the 
slope information and the correlation function of the CMB. 
We obtain more accurate results by using all multipole
moments, avoiding approximations by computing their full covariance matrix. We 
obtain the best fit for a cubic toroidal universe of scale $7200h^{-1}$Mpc 
for $n=1$. The data set a lower limit on the cell size of $4320h^{-1}$Mpc at 
95\% confidence and $5880h^{-1}$Mpc at 68\% confidence. These results 
show that the most probable cell size would be around 1.2 times larger than the 
horizon scale, implying that the 3-Torus topology is no longer an interesting 
cosmological model. 
\end{abstract}

\keywords{cosmic microwave background, large-scale structure of universe.}

\section{INTRODUCTION}

One of the basic assumptions in modern cosmology, the {\it Cosmological
Principle}, is that on large-scale average our Universe is spatially
homogeneous and isotropic. The apparent isotropy on large scales
is normally explained as a consequence of spatial homogeneity, in turn 
understood as a natural result of an ``inflationary" period of the early 
universe (see e.g. Kolb and Turner, 1990). An alternative approach to
explaining the apparent homogeneity is to assume an expanding universe with 
small and finite space sections with a non-trivial topology (Ellis and 
Schreiber, 1986), the ``small universe" model.

The ``small universe", as its name suggests, should be small enough that we 
have had time to see the universe around us many times since the decoupling
time. The topology of the spatial sections can be quite complicated 
(Ellis, 1971);
however, it is possible to obtain small universe models that reproduce a 
Friedmann-Lema\^\i tre model by choosing certain simple geometries. For
example, choosing a rectangular basic cell with sides $L_{x}$, $L_{y}$ and
$L_{z}$ and with opposite faces topologically connected, we obtain a toroidal 
topology for the small universe known as $T^{3}$. The never-ending 
repetition of this $T^{3}$ basic cell should reproduce, at least locally, the 
Friedmann-Lema\^\i tre universe model with zero curvature.
 
The small universe model has received considerable attention in the past 
few years, since the topology of the Universe is becoming an important 
problem for cosmologists. From the theoretical point of view,
it is possible to have quantum creation of the Universe with a nontrivial
topology, i.e., a multiply-connected topology (Zel'dovich and Starobinsky,
1984). From the observational side, this model has been used to explain 
``observed" periodicity in the distributions of quasars (Fang and Sato, 
1985) and galaxies (Broadhurst et al., 1990).

There are four known approaches for placing lower limits on the cell size 
of the $T^{3}$ model. The first two methods constrain the 
parameter $R$, an average length scale of the small universe, defined as 
$R=(L_{x}L_{y}L_{z})^{1/3}$. The third and fourth methods 
constrain the parameter $L/y$, the ratio between the cell size $L$, here 
defined as $L=L_{x}=L_{y}=L_{z}$, and radius of the decoupling sphere $y$, 
where $y=2cH_{o}^{-1}$. The first method constrains 
$R$ assuming that it is larger than any distinguishable structure. Using this 
method, Fairall (1985) suggests that $R > 500$Mpc. The second method 
constrains $R$ based on ``observed" periodicity in quasar redshifts. 
Attempting to identify opposite pairs of quasars, Fang and Liu 
(1988) suggested that $R > 400h^{-1}$Mpc and using quasar redshift periodicity, 
Fang and Sato (1985) suggested $R > 600h^{-1}$Mpc. The third and fourth methods 
constrain $L/y$ using the CMB. With the third method, Stevens et al. (1993)
obtain the constraint $L/y = 0.8$ using the slope information from 
the 1st year of COBE/DMR data (Smoot et al., 1992) while, with the fourth 
method, Jing and Fang (1994) obtain a best fit $L/y \approx 1.2$ using the 
correlation function from the 2-year COBE/DMR data. 

As pointed out by Zel'dovich (1973), the power spectrum of density 
perturbations is continuous (i.e., all wave numbers are possible) if the 
Universe has a Euclidean topology, and discrete (i.e., only some wave numbers 
are possible) if the topology has finite space sections. Many years later these
ideas were related with the expected CMB power spectrum (Fang and Mo, 1987; 
Sokolov, 1993; Starobinsky, 1993), mainly after the quadrupole component had 
been detected by COBE/DMR.

Our goal is to place new and accurate limits on the cell size
of a small universe using the harmonic decomposition technique to obtain the
data power spectrum (G\'{o}rski, 1994) and likelihood technique (Bunn and 
Sugiyama, 1994) to constrain $L/y$. The method that we use 
to constrain the parameter $L/y$ is quite different from previous work. The 
method adopted by Stevens et al. (1993) constrains the cell size based in 
the power spectrum of the CMB; they graphically compare the power 
spectrum of the standard model with the power spectrum expected for the small 
universe, normalizing to the $l=20$ component. Jing and Fang (1994) adopt a 
different approach: they constrain the cell size using the correlation 
function of the CMB and making the approximation that bins of the correlation 
function are uncorrelated. Our analysis, however, is exact. We compute the full 
covariance matrix for all multipole components and use this covariance matrix 
to make a $\chi^{2}$ fit of the power spectrum extracted from the 2 years 
of COBE/DMR data to the power spectrum expected for a small universe with
different cell sizes $L$. For simplicity, we limit our calculation to the 
case of a $T^{3}$ cubic universe. We present, in the next sections, a 
description of the power spectrum expected in a $T^{3}$ cubic universe model 
and the likelihood technique used to constrain $L/y$.

\section{POWER SPECTRUM OF THE $T^{3}$ UNIVERSE MODEL}

If the density fluctuations are adiabatic and the Universe is spatially
flat, the Sachs-Wolfe fluctuations in the CMB are given by
\beq{Sachs_Wolfe} \frac{\delta T}{T} (\theta,\phi) = - \frac{1}{2} 
\frac{H_{o}^{2}}{c^{2}} \sum_{\bf k} \frac{\delta_{\bf k}}{{k}^{2}} 
e^{i{\bf k} \cdot {\bf x}} \eeq
(Peebles, 1982), where ${\bf x}$ is a vector with length $y \equiv 
2cH_{o}^{-1}$ that is pointed in the direction of observation ($\theta,\phi$), 
$H_{o}$ is the Hubble constant (written here as $100h$~km/s/Mpc)
and $\delta_{\bf k}$ is the density fluctuation in Fourier space with the sum 
taken over all wave numbers {\bf k}. 

It is customary to expand the CMB anisotropy in spherical harmonics
\beq{deltaT} \frac{\delta T}{T}(\theta,\phi) = \sum_{l=0}^{\infty}
\sum_{m=-l}^{l} a_{lm} Y_{lm} (\xhat), \eeq
where $a_{lm}$ are the spherical harmonic coefficients and $\xhat$ is 
the unit vector in direction ${\bf x}$. The coefficients $a_{lm}$ are given by
\beq{alm} a_{lm} = -2\pi i^{l} \frac{H_{o}^{2}}{c^{2}} \sum_{\bf k}  
\frac{\delta_{\bf k}}{k^{2}} j_{l}(ky) Y_{lm}^{*}(\khat), \eeq
where $j_{l}$ are spherical Bessel functions of order $l$. If we assume that
the CMB anisotropy is a Gaussian random field, the coefficients $a_{lm}$ are 
independent Gaussian random variables with zero mean and variance 
\beq{almvariance} \langle | a_{lm} |^{2} \rangle = 16\pi \sum_{\bf k} 
\frac{|\delta_{\bf k}|^{2}}{(ky)^{4}}  j_{l}^{2}(ky) \eeq
(Fang and Mo, 1987; Stevens et al., 1993). Assuming a power-law power 
spectrum with shape $P(k) = |\delta_{\bf k}|^{2} = Ak^{n}$, where $A$ is the 
amplitude of scalar perturbations and $n$ the spectral index, it is possible 
to perform the sum in (\ref{almvariance}), replacing it by an integral, and to 
obtain
\beq{almpadrao} \langle | a_{lm} |^{2} \rangle = C_{2}~
\frac{\Gamma(\frac{9-n}{2})}{\Gamma(\frac{3+n}{2})}  
\frac{\Gamma(l+ \frac{n-1}{2})}{\Gamma(l+ \frac{5-n}{2})} \eeq
(see e.g. Bond and Efsthatiou, 1987). In the literature, the average over the 
canonical ensemble of universes $\langle |a_{lm}|^{2} \rangle$ is usually 
denoted by
\beq{Cl} C_{l} \equiv \langle |a_{lm}|^{2} \rangle, \eeq
where the power spectrum $C_{l}$ is related to the rms temperature fluctuation 
by $\langle | \delta T/T |_{rms}^{2} \rangle \equiv \sum_{l} (2l+1)C_{l}/ 
4\pi$. 

Note that in a Euclidean topology the Sachs-Wolfe spectrum  $C_{l}$ is an 
integral over the power spectrum; however, in the $T^{3}$ universe this is not 
the case. In this model, only wave numbers that are harmonics of the cell size 
are allowed. We have a discrete ${\bf k}$ spectrum 
\beq{kcut} {\bf k}^{2} = \sum_{i=1}^{3} \left( {\frac{2 \pi}{L_{i}}} 
\right)^2 p^{2}_{i} \eeq
(Sokolov, 1993), where $L_{1}$, $L_{2}$ and $L_{3}$ are the dimensions of the
cell and $p_{i}$ are integers. For simplicity, assuming 
$L=L_{x}=L_{y}=L_{z}$ and the same power-law power spectrum cited before, eq. 
(\ref{almvariance}) can be written as
\beq{almsu} \langle | a_{lm} |^{2} \rangle = \frac{16 \pi A}{y^{n}} 
\sum_{p_x} \sum_{p_y} \sum_{p_z} 
\left( {\frac{L}{2 \pi yp}} \right)^{4-n} j_{l}^{2} \left( {\frac{2 \pi yp}
{L}} \right), \eeq 
where $p^{2} = p_{x}^{2} + p_{y}^{2} + p_{z}^{2}$. According to 
(\ref{almsu}), the $l$th multipole of the CMB temperature is function of the 
ratio $L/y$. This shows that the more multipole components we use in our fit,
the stronger our constraints on the cell size will be. However, we cannot
use an infinite number of multipole components. The maximum number 
of multipole components, $l_{max}$, will be limited by two 
things: the limit where the map is noise dominated and the limit where we 
can truncate the Fourier series without compromising the harmonic 
decomposition technique (see G\'{o}rski, 1994).

Using eq. (\ref{almsu}), we calculated the expected power spectrum for a  
$T^{3}$ universe with different cell sizes $L/y$ from 0.1 to 3.0,
$n=1$ and $l_{max}=30$, where $l_{max}=30$ is the limit at which we truncate 
our data power spectrum. In Figure~1, we plot $l(l+1)C_{l}$ versus $l$ and 
normalize all values to the last multipole component $l=30$.
Note that for very small cells ($L \ll y$), the low order multipoles are 
suppressed. The power spectrum for small cells (as $L/y=0.1$, $0.5$ or $1.0$) 
shows the presence of ``bumps" that disappear as the cell size increases
($L/y \simgt 1.5$). The power spectrum finally becomes flat for large cell 
sizes ($L/y \simgt 3.0$). These ``bumps" can be explained if we remember
that only the harmonics of the cell size are allowed to be part of the sum in
(\ref{almsu}). When the cell size is small there are fewer modes of resonance,
and no modes larger than the cell size appear in the sum in (\ref{almsu}).
As the cell size increases, the sum approaches an integral and the $T^{3}$ 
power spectrum becomes flat.

We restrict our analysis to $n=1$. This assumption,
however, does not weaken our results, since the $T^{3}$ model with other 
$n$-values tends to fit the data as poorly as with $n=1$. For instance, we 
obtain the maximum likelihood at the same ratio $L/y$ for $n=1$ and $n=1.5$. 
This happens because the ``bumps", and not the overall slope, are responsible 
for the disagreement between the model and the data.

\section{DATA ANALYSIS}

Each DMR sky map is composed of $6144$ pixels and each pixel $i$
contains a measurement of the sky temperature at position ${\bf x}_{i}$. 
Considering that the temperatures are smoothed by the DMR 
beam and contaminated with noise, the sky temperatures are described by
\beq{deltaTobserved} \left( {\frac{\delta T}{T}} \right)_{i} = \sum_{lm} 
a_{lm} B_{l} Y_{lm}(\xhat_{i}) + n_{i}, \eeq
where $B_{l}$ is the DMR beam pattern and $n_{i}$ is the noise in pixel $i$.
We use the values of $B_{l}$ given by Wright et al. (1994a), 
which describes the actual beam pattern of the DMR horns, an imperfect gaussian 
beam. We model the quantities $n_{i}$ in (\ref{deltaTobserved}) as Gaussian 
random variables with mean $\langle n_{i} \rangle = 0$ and variance 
$\langle n_{i} n_{j} \rangle = \sigma_{i}^{2} \delta_{ij}$, assuming 
uncorrelated pixel noise (Lineweaver et al., 1994).

When we have all sky coverage, the $a_{lm}$ coefficients are given by  
\beq{almcoeff} a_{lm} = \int_{4 \pi} \left( {\frac{\delta T}{T}} \right)  
Y_{lm}^{*}(\xhat) d\Omega. \eeq
In the real sky maps, we do not have all sky coverage. Because of the
uncertainty in Galaxy emission, we are forced to remove all pixels between 
$20^{\circ}$ below and above the Galaxy plane. This cut  represents a loss of 
almost 34\% of all sky pixels and destroys the orthogonality of the spherical 
harmonics. Replacing the integral in (\ref{almcoeff}) by a sum over the number 
of pixels that remain in the sky map after the Galaxy cut, $N_{pix}$, we 
define a new set of coefficients by
\beq{blmObserved} b_{lm} \equiv w \sum_{i=1}^{N_{pix}} \left( {\frac{\delta T}
{T}} \right)_{i}
Y_{lm}^{*}(\xhat_{i}), \eeq
where the normalization is chosen to be $w \equiv 4 \pi /N_{pix}$. Substituting 
(\ref{deltaTobserved}) into (\ref{blmObserved}), we obtain 
\beq{blmFinal} b_{lm} = \sum_{l_{1}m_{1}} a_{l_{1}m_{1}} B_{l_{1}}
W_{ll_{1}mm_{1}} + w \sum_{i=1}^{N_{pix}} n_{i} 
Y_{lm}^{*}(\xhat_{i}), \eeq
with covariance 
\begin{eqnarray} \label{blmvariance} 
\langle b_{lm} b_{l'm'}^{*} \rangle & = & \sum_{l_{1}m_{1}} 
W_{ll_{1}mm_{1}}  W_{l'l_{1}m'm_{1}}  C_{l_{1}} B_{l_{1}}^{2}   \nonumber \\
& + & w^{2} \sum_{i=1}^{N_{pix}}  \sigma_{i}^{2} 
Y_{lm}^{*}(\xhat_{i}) Y_{l'm'}(\xhat_{i}), \end{eqnarray}
where
\beq{skyfactor} W_{ll_{1}mm_{1}} \equiv w \sum_{i=1}^{N_{pix}} 
Y_{lm}^{*}(\xhat_{i}) Y_{l_{1}m_{1}}(\xhat_{i}). \eeq
Defining our multipole estimates as
\beq{ClDMR} C_{l}^{DMR} \equiv \frac{1}{2l+1} \sum_{m} b_{lm} b_{lm}^{*}, \eeq
their expectation values are simply
\beq{meanClDMR} \langle C_{l}^{DMR} \rangle \equiv \frac{1}{2l+1} 
\sum_{m} \langle b_{lm} b_{lm}^{*} \rangle \eeq
and their covariance matrix $M$ is given by 
\beq{Clvariance} M_{ll'} \equiv \frac{2}{(2l+1)(2l'+1)} \sum_{mm'}
\langle b_{lm} b_{l'm'}^{*} \rangle^{2}. \eeq
The $C_{l}^{DMR}$ coefficients are not good estimates of the true 
multipole moments $C_{l}$. However, they are useful for constraining our 
cosmological parameters.

The likelihood and the $\chi^{2}$ are, respectively, defined by
%
\beq{likelihood} -2 \ln {\cal L} = \chi^{2} + \ln |{\bf M}| \eeq
and 
\beq{chi2} \chi^{2} \equiv {\bf C}^{T} {\bf M}^{-1} {\bf C}, \eeq
where ${\bf C}^{T}$ and ${\bf C}$ are $l_{max}$-dimensional row and column 
vectors with entries $C_{l} = \widehat{C}_{l}^{DMR} - \langle C_{l}^{DMR} 
\rangle$ 
and ${\bf M}$ is the covariance matrix as described in (\ref{Clvariance}) with
dimensions $l_{max}$ x $l_{max}$. Here $\widehat{C}_{l}^{DMR}$ denotes the 
$C_{l}^{DMR}$-coefficients actually extracted from the data.

Because the perturbations depend on an unknown constant $A$, the power
spectrum normalization, we have to constrain two parameters at once.
In practice, this calculation is done by fixing the ratio $L/y$ and
changing the normalization by a small factor. We multiply the first term on 
the right side of (\ref{blmvariance}) by this factor and calculate a new 
covariance matrix. Repeating this procedure for each cell size, we finally 
get a likelihood grid that constrains the ratio $L/y$ and the normalization 
parameter.

\section{RESULTS}

In Figure~2, we show the angular power spectrum $\widehat{C}_{l}^{DMR}$ 
extracted from the data. We use a 2-year combined 53 plus 90~GHz map, with 
Galaxy cut of $20^{\circ}$, monopole and dipole removed. We plot 
$l(l+1)\widehat{C}_{l}^{DMR}$ versus $l$ from 
$l=2$ to $l=30$, with bias ($\langle C_{l}^{DMR} \rangle - C_{l}$) removed
and error bars given by the diagonal terms of the 
covariance matrix $M$. In computing the bias and error bars, we assume eq.
(\ref{almpadrao}) with $n=1$. The shape of this power spectrum and its 
multipole values are consistent with values reported by Wright et al. (1994b), 
and for $l>15$ the power spectrum is basically dominated by noise.              

We computed the likelihood function ${\cal L}(L/y,\sigma_{7^{\circ}})$, using 
it to constrain the ratio $L/y$ and the normalization $\sigma_{7^{\circ}}$, 
where $\sigma_{7^{\circ}}$ is the rms variance at $7^{\circ}$. For the 
data set described above, we found the maximum likelihood at 
$(L/y,\sigma_{7^{\circ}})=(1.2,37.4 \mu K)$. In Figure~3, we plot the 
likelihood function ${\cal L}(L/y,\sigma_{7^{\circ}})$. Notice
that the likelihoods cannot be normalized because they do not converge to zero 
for very large cell sizes, i.e., the volume under the likelihood function is 
infinite. Since the likelihoods are not zero for very large cell sizes, we 
could naively consider that the probability of the universe being small  
is essentially zero. However, this conclusion is clearly 
exaggerated and based on the fact that we multiplied our likelihoods by a
uniform prior, and there is nothing special about adopting a uniform 
prior. In order to obtain rigorous confidence limits for our analysis, we 
replace the maximum likelihood fit by a minimum $\chi^{2}$ fit.

We compute the chi-squared function $\chi^{2}(L/y,\sigma_{7^{\circ}})$ and use
it to constrain the ratio $L/y$ and the normalization $\sigma_{7^{\circ}}$. 
In Figure~4, we plot the probability that the $T^{3}$ model is consistent with 
the data as a function of the ratio $L/y$ and the normalization 
$\sigma_{7^{\circ}}$ (bottom). Confidence limits of 68\%, 95\% and 99.7\% are
shown in the contour plot (top). We found the
highest consistency probability (minimum $\chi^{2}$) at 
$(L/y,\sigma_{7^{\circ}}) = (1.2,49.7 \mu K)$, represented by a cross in the 
contour plot. Removing the quadrupole, we obtained similar 
results, see Table~1 for the lower limits on cell sizes. We obtain the
constraint $L/y = 1.2_{-0.48}^{+ \infty}$ at 95\% confidence.  We cannot place 
an upper limit on the cell size: all large cells are equally probable.

\section{CONCLUSIONS}

The strong constraint from our analysis comes from the predicted power 
spectrum of the $T^{3}$ universe; see Figure~1. According to this plot, a 
reduction in the cell size to values below the horizon scale should 
suppress the quadrupole and low multipole anisotropies, while the suppression 
is negligible if the cell is very large, at least, larger than the horizon.
It is possible to notice these properties in Figures~3 and~4: 
both favor large cell sizes. The observed presence of the quadrupole 
and other low order anisotropies automatically constrains our cell to be very 
large. In other words, even before making the $\chi^{2}$ fit, we expect to 
obtain very large cells.

We remind the reader that our analysis is for $n=1$. We made this assumption 
because the results of fitting the $T^{3}$ model seem to be relatively 
insensitive to
changes in $n$ and the ``bumps", not the overall slope, are responsible for the 
poor fit between the model and the data. In other words, our results are
independent of any particular inflationary model.

From the COBE/DMR data, we obtain the best $\chi^{2}$ fit for a toroidal 
universe with $L/y=1.2$, which corresponds to a cell size of 
$L=7200 h^{-1}$Mpc. A cell size below 72\% of the size of the horizon 
($L/y < 0.72$) is incompatible with the COBE measurements at 95\% confidence, 
and a cell size below the size of the horizon ($L/y < 0.98$) is ruled out at 
68\% confidence. Since the $T^{3}$ topology is interesting if the cell size is 
considerably smaller than the horizon, this model loses most of its appeal.

\bigskip

We would like to thank Jon Aymon, Douglas Scott, Joseph Silk, Daniel 
Stevens and Max Tegmark for many 
useful comments and help with the manuscript. AC acknowledges SCT-PR/CNPq 
Conselho Nacional de Desenvolvimento Cient\'\i fico e Tecnol\'{o}gico for her 
financial support under process No.201330/93-8(NV). This work was supported in 
part by the Director, Office of Energy Research, Office of High Energy and 
Nuclear Physics, Division of High Energy Physics of the U.S. Department of 
Energy under contract No.DE-AC03-76SF00098.

\clearpage

\begin{table}
\caption{Lower limits on L/y} 
\begin{center}
\begin{tabular}{ccc}
\hline
\multicolumn{1}{c}{Confidence Level}  &
\multicolumn{1}{c}{L/y with $C_{2}$}    &
\multicolumn{1}{c}{L/y without $C_{2}$} \\ 
\hline
68  \% &0.98 &0.97 \\
90  \% &0.75 &0.68 \\
95  \% &0.72 &0.65 \\ 
99.7\% &0.61 &0.60 \\
\hline
\end{tabular}
\end{center}
\end{table} 

\clearpage

\centerline{\bf REFERENCES}
\bigskip
\noindent
Bond, J.R. \& Efsthatiou, G. 1987, Mon. Not. R. Astr. Soc., 226:655.\\
Broadhurst, T.J. et al. 1990, Nature, 343:726.\\
Bunn, E. \& Sugiyama, N. 1994, preprint (astro-ph/9407069). \\
Ellis, G.F.R. 1971, Gen. Rel. and Grav., 2(1):7.\\
Ellis, G.F.R. \& Schreiber, G. 1986, Phys. Lett. A, 115(3):97.\\
Fairall, A.P. 1985, Mon. Not. R. Astr. Soc. of South. Africa, 44(11):114.\\
Fang, L.Z. \& Liu, Y.L. 1988, Mod. Phys. Lett. A, 3(13):1221.\\
Fang, L.Z. \& Mo, H. 1987, Mod. Phys. Lett. A, 2(4):229.\\
Fang, L.Z. \& Sato, H. 1985, Gen. Rel. and Grav., 17(11):1117.\\
G\'{o}rski, K.M. 1994, Ap. J. Lett., 430:L85.\\
Jing, Y.P. \& Fang, L.Z., 1994, Phy. Rev. Lett., 73(14):1882.\\
Kolb, E.W. \& Turner, M.S. 1990, The Early Universe, Addison-Wesley.\\
Lineweaver, C. et al. 1994, Ap. J., 436:452. \\
Peebles, P.J.E. 1982, Ap. J. Lett., 263:L1. \\
Smoot, G.F. et al. 1992, Ap. J. Lett., 396:L1.\\
Sokolov, I.Y. 1993, JETP Lett., 57(10):617.\\
Starobinsky, A.A. 1993, JETP Lett., 57(10):622.\\
Stevens, D. et al. 1993, Phys. Rev. Lett., 71(1):20.\\
Wright, E.L. et al. 1994a, Ap. J., 420:1.\\
Wright, E.L. et al. 1994b, Ap. J., 436:433. \\
Zel'dovich, Ya B. 1973, Comm. Astrophys. Space Sci., 5(6):169.\\
Zel'dovich, Ya B. and Starobinsky, A.A. 1984, Sov. Astron. Lett., 10(3):135.\\

\clearpage

\centerline{\bf FIGURE CAPTIONS}
\bigskip
\noindent
Figure~1: Expected power spectrum for the $T^{3}$ universe model 
with $n=1$ for different cell sizes with $L/y$ from 0.1 to 3.0.\\
Figure~2: Power spectrum of the 2-year combined 53+90~GHz 
COBE/DMR data with 
bias removed and the error bars given by the diagonal terms of the covariance 
matrix $M$.\\
Figure~3: The likelihood function ${\cal L}(L/y,\sigma_{7^{\circ}})$ 
for the $T^{3}$ universe model with $n=1$. \\
Figure~4: The probability that the $T^{3}$ model is consistent with 
the data is 
plotted as a function of the ratio $L/y$ and the normalization 
$\sigma_{7^{\circ}}$ (bottom). Confidence limits of 68\%, 95\% and 99.7\% are
shown in the contour plot (top). We found the highest consistency probability
(minimum $\chi^{2}$) at $L/y=1.2$, represented by a cross in the contour plot.

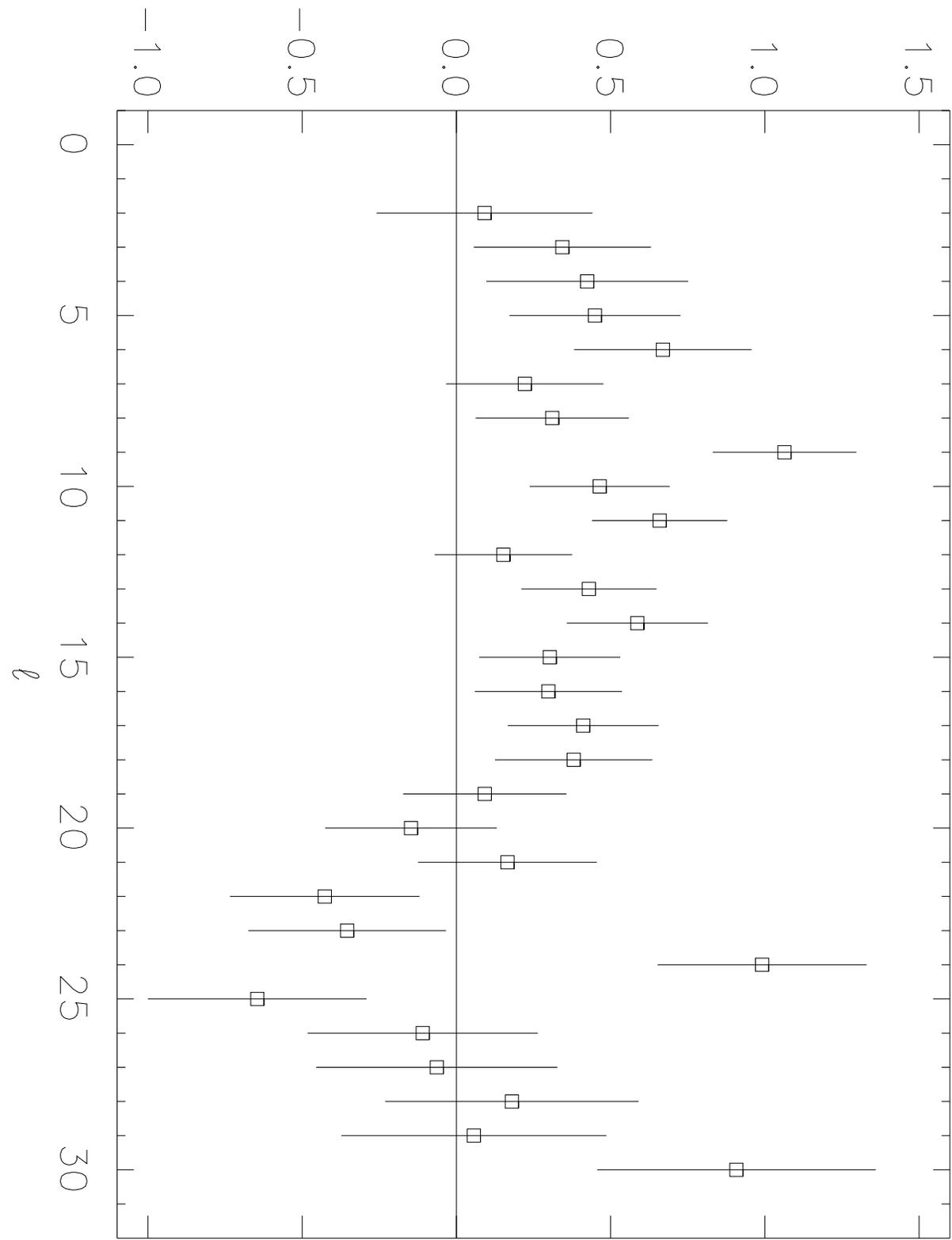
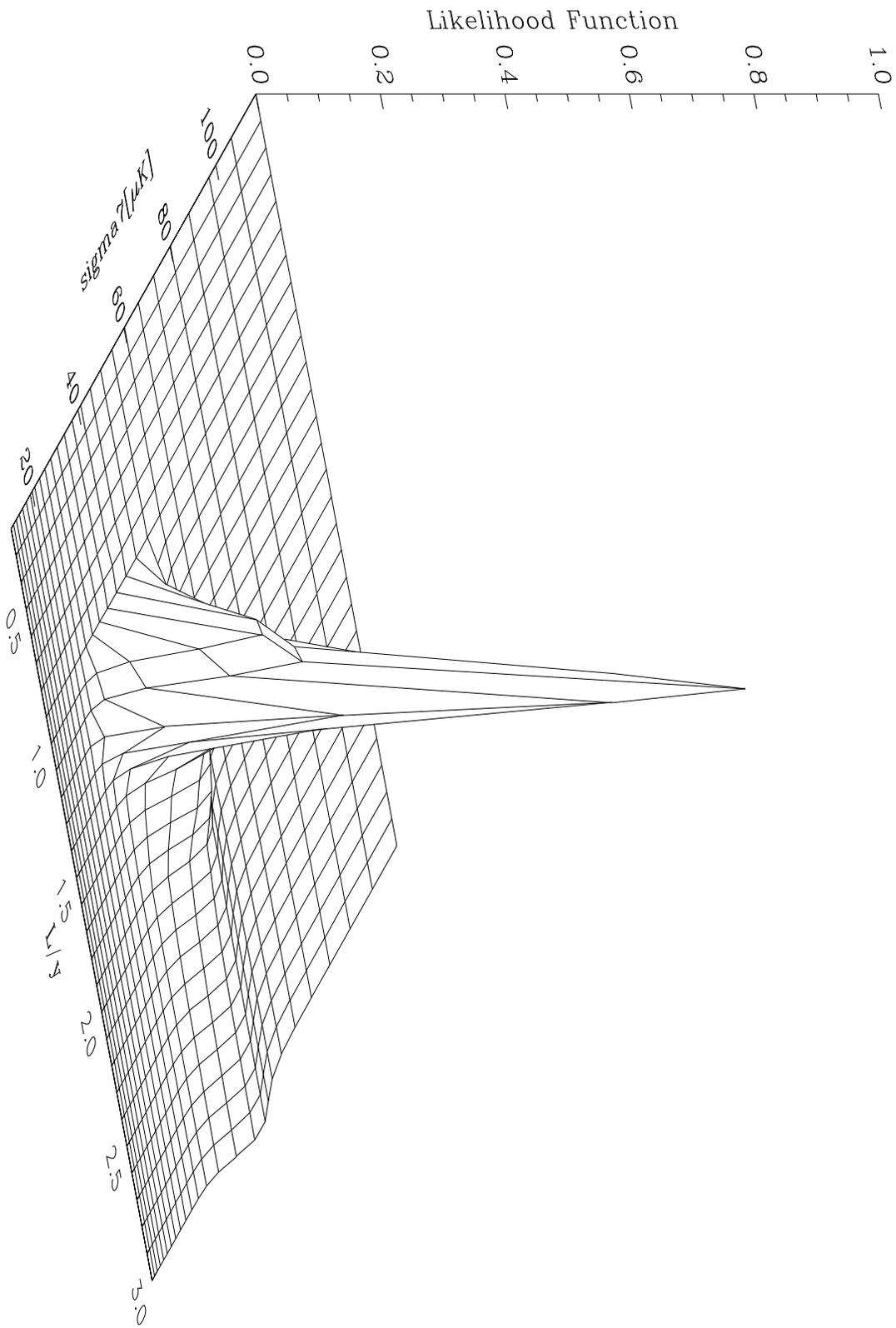
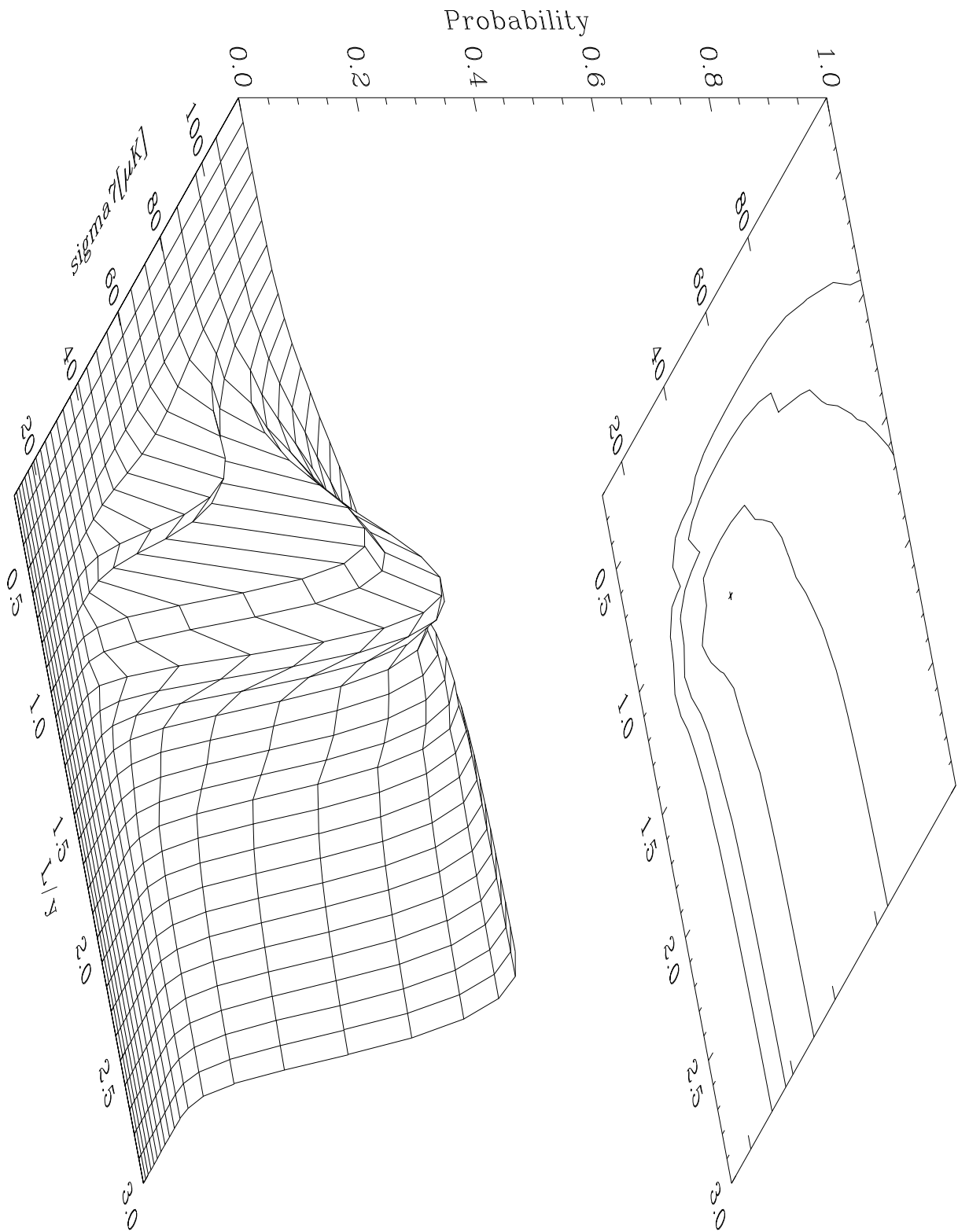
\end{document}